\documentstyle[preprint,aps,epsf]{revtex}
\def\bea{\begin{eqnarray}}
\def\beann{\begin{eqnarray*}}
\def\beq{\begin{equation}}
\def\eea{\end{eqnarray}}
\def\eeann{\end{eqnarray*}}
\def\eeq{\end{equation}}
\def\nn{\nonumber}

\def\ran{\rangle}
\def\lan{\langle}

\newcommand{\bx}{\bbox{x}}
\newcommand{\by}{\bbox{y}}

\newcommand{\bcdot}{\bbox{\cdot}}

\newcommand{\bk}{\bbox{k}}
\newcommand{\bp}{\bbox{p}}
\newcommand{\bq}{\bbox{q}}

\newcommand{\bS}{\bbox{S}}

\newcommand{\balpha}{\bbox{\alpha}}

\newcommand{\bsigma}{\bbox{\sigma}}
\newcommand{\bnabla}{\bbox{\nabla}}

\begin{document}
\title{Chiral corrections to baryon properties with composite pions}
\author{P.J.A. Bicudo}
\address{Departamento de F\'{\i}sica and CFIF-Edif\'{\i}cio Ci\^encia, 
Instituto Superior T\'ecnico\\
Avenida Rovisco Pais, 1096 Lisboa Codex, Portugal}
\author{G. Krein}
\address{Instituto de F\'{\i}sica Te\'orica, Universidade Estadual Paulista \\
Rua Pamplona, 145 - 01405-900 S\~ao Paulo, SP,  Brazil}
\author{J.E.F.T. Ribeiro}
\address{Departamento de F\'{\i}sica and CFIF-Edif\'{\i}cio Ci\^encia, 
Instituto Superior T\'ecnico\\
Avenida Rovisco Pais, 1096 Lisboa Codex, Portugal}

\maketitle

\begin{abstract}
A calculational scheme is developed to evaluate chiral corrections to 
properties of composite baryons with composite pions. The composite baryons 
and pions are bound states derived from a microscopic chiral quark model. 
The model is amenable to standard many-body techniques such as the BCS and 
RPA formalisms. An effective chiral model involving only hadronic degrees 
of freedom is derived from the macroscopic quark model by projection onto 
hadron states. Chiral loops are calculated using the effective 
hadronic Hamiltonian. A simple microscopic confining interaction is used 
to illustrate the derivation of the pion-nucleon form factor and the 
calculation of pionic self-energy corrections to the nucleon and 
$\Delta(1232)$ masses.
\end{abstract}
\date{ }
\pacs{13.75.Gx, 12.39.Fe, 12.39.Ki, 24.85.+p}
\narrowtext

\section{Introduction} 
The incorporation of chiral symmetry in quark models is an important issue 
in hadronic physics. The subject dates back to the early works
~\cite{{CT},{Bar},{BR},{Vento},{Jaffe},{CBMorig}} aimed at restoring 
chiral symmetry to the MIT bag model~\cite{MIT}. The early attempts were 
based on coupling elementary pion fields directly to quarks. A great variety 
of chiral quark-pion models have been constructed since then and the subject 
continues to be of interest in the recent literature~\cite{{GR},{TK},{Isg}}. 
Despite the long history, there are many important open questions in 
this field. In the present paper, we are concerned with one of such 
questions, namely the coupling of the pion as a quark-antiquark bound state 
to the baryons. Starting from a model chiral quark Hamiltonian, we construct 
an effective low-energy chiral pion-baryon Hamiltonian appropriate for 
calculating chiral loop corrections to hadron properties. The composite 
pion and baryon states are determined by the same underlying quark chiral 
dynamics. 

The model we use belongs to a class of quark models  inspired in the 
Coulomb gauge QCD Hamiltonian~\cite{Coul1st} and generalizes the 
Nambu--Jona-Lasinio model~\cite{NJL} to include confinement and asymptotic 
freedom. This class of 
models is amenable to standard many-body techniques such as the BCS formalism 
of superconductivity. The initial studies within these models were aimed at 
studying the interplay between confinement and dynamical chiral symmetry 
breaking (D$\chi$SB), and concentrated on critical couplings~\cite{LeYaou-pot}
for D$\chi$SB and light-meson spectroscopy~\cite{LeYaou-mes}. The model has 
been extended to study the pion beyond BCS level and meson resonant decays in 
the context of a generalized resonating group method~\cite{BicRib1}. 
Since the model is formulated on the basis of a Hamiltonian, it provides a 
natural way to study finite temperature and chemical potential quark 
matter~\cite{Mishra}. The model and the many-body techniques to solve 
it make direct contact with first-principle developments such as 
nonperturbative renormalization-group treatments of the QCD 
Hamiltonian~\cite{renorm} and Hamiltonian lattice QCD~\cite{HLQCD}.

One important development of the model, in the context of the present paper, 
was its extension in Ref.~\cite{baryons} to baryon structure. 
In Ref.~\cite{baryons}
a variational calculation was implemented for the masses of baryons and it was
shown that a sizeable $\Delta(1232)$-$N$ mass difference is obtained from the 
same underlying hyperfine interaction that gives a reasonable value for 
the $\pi$-$\rho$ mass difference. This hyperfine interaction, along with 
other spin-dependent interactions like tensor and spin-orbit, stem from 
Bogoliubov-Valatin rotated spinors that depend on the ``chiral angle". The
chiral angle gives the extent of the chiral condensation in the vacuum and
determines the chiral condensate. The very same variational 
wave function was used later for studying $S$-wave kaon-nucleon~\cite{KN} 
scattering and the repulsive core of the nucleon-nucleon force~\cite{NN}. 
Both calculations obtain $S$-wave phase-shifts that compare reasonably well 
with experimental data. A remarkable feature of all these results for the 
low-lying spectrum of mesons and baryons and $S$-wave scattering phase-shifts 
is that they are obtained with a single free parameter, the strength of the 
confining potential. 

In the present paper we go one step forward in the development of the model 
and set up a calculational scheme to treat chiral corrections in hadron 
spectroscopy. In a recent publication~\cite{fpiNN}, two of us 
have calculated the pion-nucleon coupling constant in this model and obtained 
reasonable agreement with its experimental value. Here, we are interested in 
developing a scheme for calculating chiral corrections to hadron properties.
We study the requirements to obtain in the context of the model the correct 
leading nonanalytic behavior (LNA) of chiral loops. Our scheme follows the 
standard practice~\cite{CBMorig}~\cite{CBMrev} of constructing an effective 
baryon-pion Hamiltonian by projecting the quark Hamiltonian onto a 
Fock-space basis of single composite hadronic states. Chiral loop 
corrections are then calculated with the effective Hamiltonian in 
time-ordered perturbation theory. The difference here is that while in the
previous works the pion is an elementary particle, in our approach the pions
are composites described by a Salpeter amplitude. 

A difficulty appears in the implementation of the projection of the 
microscopic quark Hamiltonian onto the composite hadron states, which is 
not present when the pion is treated as an elementary particle. The 
difficulty is related to the two-component nature of the Salpeter amplitude 
of the pion. The two components correspond to positive and negative energies
(forward and backward moving, in the language of of time-ordered perturbation 
theory), they are 2 $\times$ 2 matrices in spin space and are called 
energy-spin (E-spin) wave functions. For the pion, the negative-energy 
component is as important as the positive-energy one -- in the chiral 
limit they are equal -- because of the Goldstone-boson nature of the pion. 
Because of this, the Fock-space representation of the pion state is not 
simple. We overcome the difficulty by rephrasing the formalism of the 
Salpeter equation in terms of the RPA (random-phase approximation) equations 
of many-body theory. The single-pion state is obtained in terms of a 
creation operator acting on the RPA vacuum. The pion creation operator is a 
linear combination of creation and annihilation operators of pairs of 
quark-antiquark operators; the positive-energy Salpeter component comes 
with the creation operator of the quark-antiquark pair, and the 
negative-energy one comes with the annihilation operator of the 
quark-antiquark pair. In this way, the projection of the microscopic 
quark Hamiltonian onto the single hadron states becomes feasible and simple.

The paper is organized as follows. In the next Section we review the basic 
equations of the model. We show the relationship of the formalisms of the 
Salpeter equation and of the many-body technique of the RPA.  
In Section~\ref{sec:vertex} we derive the pion-baryon vertex function in terms
of the bound-state Salpeter amplitudes for the pion and the baryon. We obtain 
an expression that is valid for a general microscopic quark interaction, not 
restricted to a specific form of the potential. Given the pion-baryon vertex 
function, we derive the expression for the baryon self-energy correction in 
Section~\ref{sec:selfenergy}.  Although the derivation of the expression for 
the self-energy is well-known in the literature, we repeat it here to make 
the paper easier to read. In Section~\ref{sec:numerical} we obtain numerical 
results for the pion-baryon form factor and coupling constants. Numerical 
results and the discussion of the LNA contributions to the baryon masses are 
presented in Section~\ref{sec:self-LNA}. Section~\ref{sec:conclusion} presents
our conclusions and future directions.

\section{The model}
\label{sec:model}

The Hamiltonian of the model is of the general form
\beq
H = H_0 + H_I, 
\label{H}
\eeq
where $H_0$ is the Dirac Hamiltonian
\beq
H_0 = \int d\bx \, \psi^{\dagger}(\bx)\, (-i \balpha\bcdot\bnabla + 
\beta m_q )\,\psi(\bx),
\label{H0}
\eeq
with $\psi(\bx)$ the Dirac field operator, and $H_I$ a chirally 
symmetric four-fermion interaction 
\beq
H_I = \frac{1}{2} \int d\bx \int d\by \, \psi^{\dagger}(\bx)T^a\Gamma
\psi(\bx) \,V_{\Gamma}(\bx-\by)\,\psi^{\dagger}(\by)T^a\Gamma\psi(\by).
\label{HI}
\eeq
Here, $T^a=1/2\,\lambda^a, a=1,\dots,8$ are the generators of the color
SU(3) group, $\Gamma$ is one, or a combination of Dirac matrices, and 
$V_{\Gamma}$ contains a confining interaction and other spin-dependent 
interactions. One example of $V_{\Gamma}$ will be presented in 
section~\ref{sec:numerical}, when we make a numerical application of the
formalism. 

Once the model Hamiltonian is specified, the next step consists in 
constructing an explicit but approximate vacuum state of the Hamiltonian 
in the form of a pairing ansatz. This is most easily implemented in the
form of a Bogoliubov-Valatin transformation (BVT). The transformation 
depends on a pairing function, or chiral angle $\varphi$ that determines
the strength of the pairing in the vacuum. The quark field operator is 
expanded as
\beq
\psi (\bx)  = \int \frac{d\bq}{(2\pi)^{3/2}} \left[ u_s({\bq})\,
b(\bq) + v_s(\bq)\,d^{\dag} (-\bq)\right] e^{i\bq\cdot\bx} \,,
\end{equation}
where the quark and antiquark annihilation operators $b$ and $d$ annihilate 
the paired vacuum, or BCS state, $|0_{\rm BCS}\ran$. Here the spinors 
$u_s(\bq)$ and $v_s(\bq)$ depend upon the chiral angle $\varphi$ as
\bea
u_s({\bq}) &=& \frac{1}{\sqrt{2}}\left\{ [1+\sin \varphi (q)]^{\frac{1}{2}} 
+ [1-\sin \varphi (q)]^{\frac{1}{2}}\balpha\cdot\hat{\bq}\right\} 
u_{s}^0 \,,\\
v_s({\bq}) &=& \frac{1}{\sqrt{2}}\left\{ [1+\sin \varphi (q)]^{\frac{1}{2}}
- [1-\sin \varphi (q)]^{\frac{1}{2}}\balpha\cdot\hat{\bq}\right\} 
v_{s}^0\,,
\label{u_and_v}
\eea
where $u_{s}^0$ and $v_{s}^0$ are the spinor eigenvectors of Dirac matrix 
$\gamma^0=\beta$ with eigenvalues $\pm 1$, respectively.

The chiral angle can be determined from the minimization of the vacuum
energy density,
\bea
\frac{E_{vac}}{V} &=& \int \frac{d\bq}{(2\pi)^3}{\rm Tr}\left[
\balpha\cdot\bq\,\Lambda^-(\bq)\right] \nn\\
&+& \frac{1}{2}
\int \frac{d\bq}{(2\pi)^3}\frac{d\bq'}{(2\pi)^3}\tilde V(\bq-\bq')
{\rm Tr}\left[T^a\Gamma \Lambda^+(\bq) T^a\Gamma \Lambda^-(\bq')\right],
\eea
where ${\rm Tr}$ is the trace over color, flavor, and Dirac indices,
and
\beq
\Lambda^+(\bq) = \sum_s u_s(\bq) u^{\dagger}_s(\bq)\,,\hspace{1.0cm}
\Lambda^-(\bq) = \sum_s v_s(\bq) v^{\dagger}_s((\bq) .  
\eeq
The minimization of the vacuum energy leads to the gap equation,
\beq
A(q)\cos \varphi(q) - B(q)\sin \varphi(q) =0,
\label{gap}
\eeq
with
\bea
A(q) &=& m + \frac{1}{2} \int \frac{d\bq'}{(2\pi)^3}\, 
\tilde V_{\Gamma}(\bq-\bq') \, {\rm Tr}\left[\beta\,T^a\Gamma
\left(\Lambda^+(\bq')-\Lambda^-(\bq)\right)T^a\Gamma\right], \\
B(q) &=& q + \frac{1}{2} \int \frac{d\bq'}{(2\pi)^3}\, 
\tilde V_{\Gamma}(\bq-\bq') {\rm Tr}\left[\balpha\cdot\hat{\bq}\, 
T^a\Gamma\left(\Lambda^+(\bq')-\Lambda^-(\bq)\right)T^a\Gamma\right],
\eea
where $\tilde V_{\Gamma}(\bq)$ is the Fourier transform of $V_{\Gamma}(\bx)$,
\beq
\tilde V_{\Gamma}(\bq) = \int d\bx\, e^{i\bq\cdot\bx}\,V_{\Gamma}(\bx).
\eeq

The pion bound-state equation is given by the field-theoretic Salpeter 
equation. The wave function has two components, $\phi^+$ and $\phi^-$, the 
positive- and negative-energy components. Each of the $\phi$'s is a 
$2~\times~2$ matrix in spin space, $\phi^+_{s_1,s_2}$ and $\phi^-_{s_1,s_2}$. 
For this reason, the $\phi$'s are also called energy-spin (E-spin) 
wave functions. The $\phi$'s satisfy the following coupled integral equations:
\bea
&& [M(\bq)-E(\bq_+)-E(\bq_-)]\phi^+_{\bk}(\bq) = 
u^{\dagger}(\bq_+)\,K_{\phi}(\bk,\bq)\,v(\bq_-) \label{phi+} , \\ 
&&[M(\bq)+E(\bq_+)+E(\bq_-)]\phi^{-T}_{\bk}(\bq) =
v^{\dagger}(\bq_+)\,K_{\phi}(\bk,\bq)\,u(\bq_-) ,
\label{phi-}
\eea
where $\bq_{\pm} = \bq \pm \bk/2$, and the kernel $K_{\phi}(\bk,\bq)$ is 
given by 
\bea
K_{\phi}(\bk,\bq) &=& \int \frac{d\bq'}{(2\pi)^3} \tilde V_{\Gamma}(\bq-\bq')\,
T^a \Gamma \,\bigl[u(\bq'_+)\,\phi^+_{\bk}(\bq')\,v^{\dagger}(\bk'_-) \nn\\
&+& v(\bk'_+)\,\phi^{-T}_{\bk}(\bq')\,u^{\dagger}(\bq'_-)\bigr]\,T^a\Gamma .
\label{Kphi}
\eea
Here, the super-script $T$ on $\phi^{-T}$ means spin-transpose of $\phi^{-}$, 
$\phi^{-T}_{s_1,s_2}=\phi^{-}_{s_2,s_1}$. The amplitudes are normalized as
\beq
\int \frac{d\bq}{(2\pi)^3} \left[\phi^{+*}_{\bk}(\bq) \phi^{+}_{\bk'}(\bq)
- \phi^{-*}_{\bk}(\bq) \phi^{-}_{\bk'}(\bq)\right] = \delta(\bk - \bk') \,.
\label{norm-phi}
\eeq

In the language of many-body theory, these equations can be identified with 
the RPA equations~\cite{LECot}. In the RPA formulation, one writes for the 
one-pion state (suppressing isospin quantum numbers)
\beq
|\pi(\bk)\ran = M^{\dagger}_{\pi}(\bk)|0_{\rm RPA}\ran ,
\label{1pi}
\eeq
where $|0_{\rm RPA}\ran$ is the RPA vacuum, which contains correlations 
beyond the BCS pairing vacuum $|0_{\rm BCS}\ran$, and $M^{\dagger}_{\pi}(\bk)$
is the pion creation operator
\beq
M^{\dagger}_{\pi}(\bk) = \int \frac{d\bq}{(2\pi)^{3/2}} 
\left[\phi^+_{\bk}(\bq)\, b^{\dagger}(\bq_+)\, d^{\dagger}(\bq_-) - 
\phi^-_{\bk}(\bq)\,b(\bq_+) d(\bq_-)\right] ,
\label{Mop}
\eeq
where the $\bq_{\pm}$ were defined above. In this formulation, 
Eqs.~(\ref{phi+}) and (\ref{phi-}) above are obtained from the RPA equation 
of motion:
\beq
\lan \pi(\bk)| [H,M^{\dagger}_{\pi}]|0_{\rm RPA}\ran = 
(E_{\pi}(\bk)-E_{vac})\, \lan \pi(\bk)|M^{\dagger}|0_{\rm RPA}\ran .
\label{RPAeqs}
\eeq
The normalization is such that
\beq
\lan \pi(\bk)|\pi(\bk')\ran =
\lan 0_{\rm RPA}|[M_{\pi}(\bk),M^{\dagger}_{\pi}(\bk')]|0_{\rm RPA}\ran
= \delta(\bk-\bk')\,.
\label{normM}
\eeq

The verification of D$\chi$SB consists in finding nontrivial solutions to the 
gap equation, Eq.~(\ref{gap}), and the existence of solutions for the pion 
wave function. Refs.~\cite{Coul1st} and \cite{BicRib1} show that for a 
confining force, there is always a nontrivial solution for the gap and pion 
Salpeter equations. Moreover, good numerical values are obtained for the 
chiral parameters such as the pion decay constant and the chiral condensate 
when an appropriate spin-dependent potential is used~\cite{Bic}. 

The inclusion of RPA correlations beyond BCS pairing was shown in 
Ref~\cite{LECot} to have a dramatic effect on the mass spectrum of the 
pseudo scalar mesons ($\pi$ and $\eta$), while it has almost no effect on 
the mass spectrum of vector mesons ($\rho$ and $\omega$). One can trace this 
effect to the fact that the pseudo scalar mesons have a sizeable ``negative 
energy'' component wave function, while the vector mesons have a very 
small negative energy component~\cite{BicRib1}. For baryons (such as 
nucleon and $\Delta$), since they do not have a sizeable negative energy 
component~\cite{baryons}, one expects that they can be reliably obtained 
from the BCS vacuum. We write therefore for the one-baryon state
\beq
|B^{(0)}(\bp) \ran = B^{(0)\dagger}(\bp)|0_{\rm BCS}\ran ,
\label{1bar}
\eeq
where the baryon creation operator $B^{(0)\dagger}(\bp)$ is given by
\bea
B^{(0)\dagger}(\bp) &=& \int d\bq_1\,d\bq_1\,d\bq_1\, 
\delta(\bp-\bq_1-\bq_2-\bq_3)\,\Psi_{\bp}(\bq_1\bq_2\bq_3)\,
\epsilon^{c_1c_2c_3}\,\chi^{f_1f_2f_3s_1s_2s_3}\,\nn\\
&\times& b^{\dagger}_{c_1s_1f_1}(\bq_1)\,b^{\dagger}_{c_2s_2f_2}(\bq_2)\,
b^{\dagger}_{c_3s_3f_3}(\bq_3)\,.
\label{Bop}
\eea
Here $\epsilon^{c_1c_2c_3}$ is the Levi-Civita tensor, which guarantees
that the baryon is a color singlet and $\chi^{f_1f_2f_3s_1s_2s_3}$ are the 
spin-isospin coefficients. The wave function $\Psi_{\bp}(\bq_1\bq_2\bq_3)$
is determined variationally~\cite{baryons}. The index $(0)$ on the baryon
operators indicates a bare baryon, i.e. a baryon without pion cloud 
corrections. 

The important fact to notice here is that the baryon wave function depends
on the chiral angle $\varphi$ and as such, spin splittings and other 
properties are determined by the same physics that determines vacuum
properties. The pion-baryon vertex, that we will discuss in the next 
section, will therefore depend on the chiral angle not only because
of the pion, but also because of the baryon wave function. Numerical 
results for the masses of the the nucleon and $\Delta(1232)$ have been 
obtained previously~\cite{baryons} and are of the right order of magnitude 
as compared with experimental values. Of course, fine-tuning with different 
spin-dependent interactions can improve the numerical values of the 
calculated quantities. 

\section{Pion-baryon vertex function}
\label{sec:vertex}

We obtain an effective baryon-pion Hamiltonian by projecting the model
quark Hamiltonian onto the one-pion and one-baryon states, Eqs.~(\ref{1pi}) 
and (\ref{1bar}). We use a shorthand notation. For the bare baryons, i.e.
baryons without pionic corrections, we use the indices $\alpha, \beta, 
\cdots$ to indicate all the quantum numbers necessary to specify 
the baryon state, such as spin, flavor and center-of-mass momentum. 
For the pion, we use $j, k, \cdots$ to specify all the quantum numbers 
of the pion state. With this notation, the effective Hamiltonian 
can be obtained as
\bea
H &=& \sum_{\alpha\beta} |\alpha\ran \lan\alpha|H|\beta\ran \lan \beta| 
+ \sum_{j k}|j\ran \lan j|H|k\ran \lan k| \nn \\
&+& \sum_{j \alpha\beta}\Bigl( 
|\alpha\ran|j\ran \lan\alpha,j|H_I|\beta\ran \lan \beta|
+ |\beta\ran \lan \beta|H_I|j,\alpha\ran \lan\alpha| \lan j| 
\Bigr) \,.
\label{Hproj}
\eea
This leads to an effective Hamiltonian that can be written as the sum of the 
single-baryon and single-pion contributions, and the pion-baryon vertex:
\beq
H = H_0 + W,
\label{partit}
\eeq
where $H_0 = H_B + H_{\pi}$ contains the single-baryon and single-pion
contributions 
\beq
H_B = \sum_{\alpha} E^{(0)}_{\alpha} \, B^{(0)\dag}_{\alpha}\, 
B^{(0)}_{\alpha}, \hspace{1.0cm} 
H_\pi = \sum_j E_j \, M^{\dagger}_j \, M_j,
\label{H0-Hpi} 
\eeq
and $W$ is the pion-baryon vertex
\beq
W = \sum_{j \alpha\beta}W^j_{\alpha\beta} \,B^{(0)\dag}_{\beta} \, 
B^{(0)}_{\alpha} \, M_j  + {\rm h.c.} \,.
\label{W}
\eeq
Here, $B^{(0)\dag}_{\alpha}$ and $B^{(0)}_{\alpha}$ ($M^{\dagger}_j$ and 
$M_j$) are the baryon (pion) creation and annihilation operators, discussed 
in the previous section. Note that we have assumed that states with different 
quantum numbers are orthogonal. Note also that in writing the state
$|j,\alpha\ran$ we have implicitly assumed that the negative-energy component
of the baryon is negligible and the baryon creation operator acting on the 
RPA vacuum has the same effect as acting on the BCS vacuum. 

We note that the projection of the microscopic quark Hamiltonian to an 
effective hadronic Hamiltonian can be obtained in a systematic and controlled 
way using a mapping procedure~\cite{mapping}. We do not follow such a procedure
here because we are mainly concerned with tree-level pion-baryon 
coupling and the projection we are using is enough to obtain the desired 
effective coupling. For processes that involve quark-exchange, such as 
baryon-meson or baryon-baryon interactions, a mapping procedure would be
useful. In a future publication, we intend to address such processes in the
context of the present model.

The single-hadron Hamiltonians $H_0$ and $H_{\pi}$ give
\bea
H_0 |B^{(0)}_{\alpha}\ran = E^{(0)}_B |B^{(0)} \ran \,,
\hspace{1.0cm}H_{\pi} |\pi \ran = E^{(0)}_{\pi} |\pi\ran \,,
\label{H0Bpi}
\eea
with $E^{(0)}_B  = M^{(0)}_B$ and  $E^{(0)}_{\pi} = m^{(0)}_\pi$ in the
rest-frame. The vertex $W$ gives the coupling of the pion to the baryon and, 
as we will show later, leads to loop corrections to the baryon self-energy. 

\begin{figure}
\centerline{
\hspace{3.0cm}{\epsfxsize=15.0cm\epsfbox{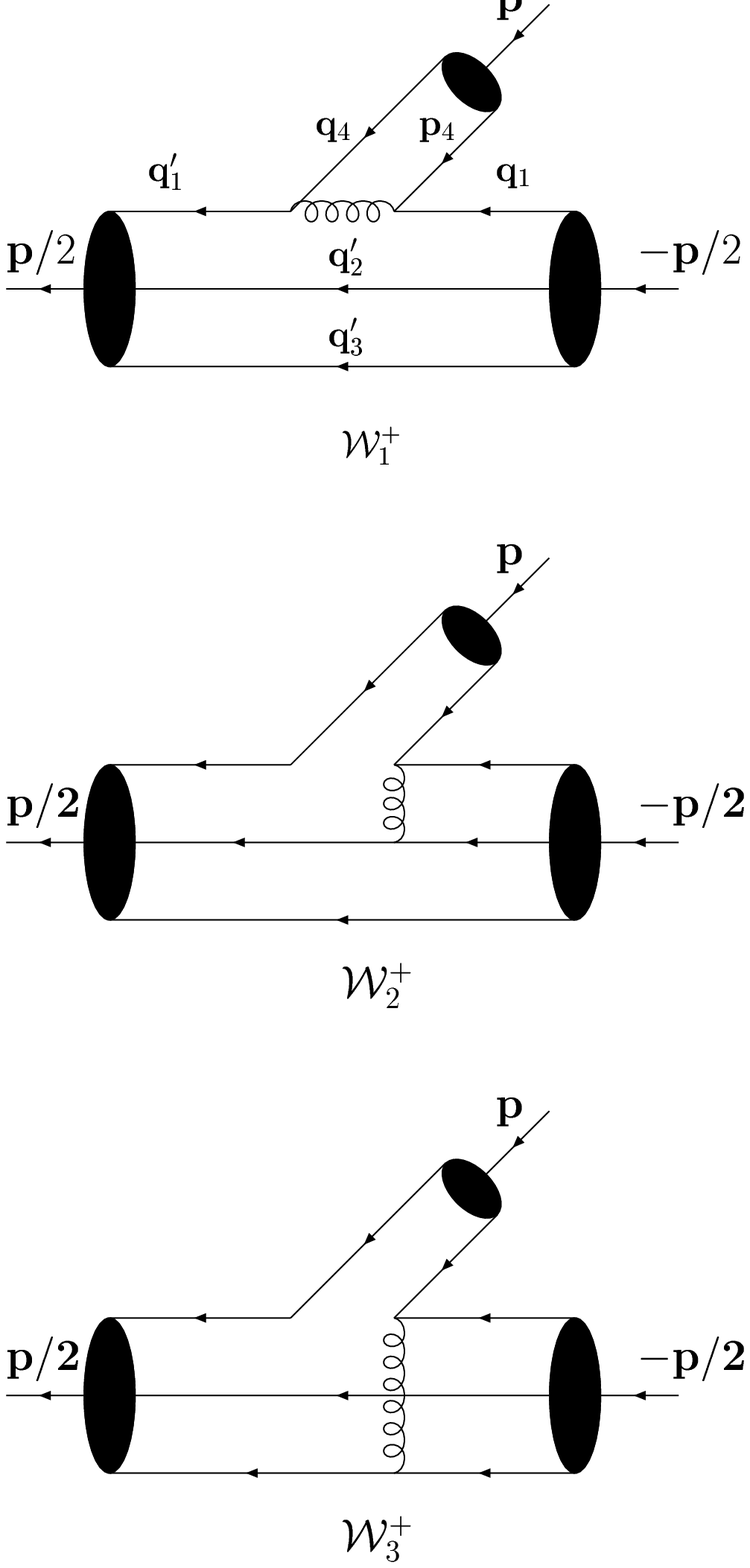}}\hspace{-6.0cm}
{\epsfxsize=15.0cm\epsfbox{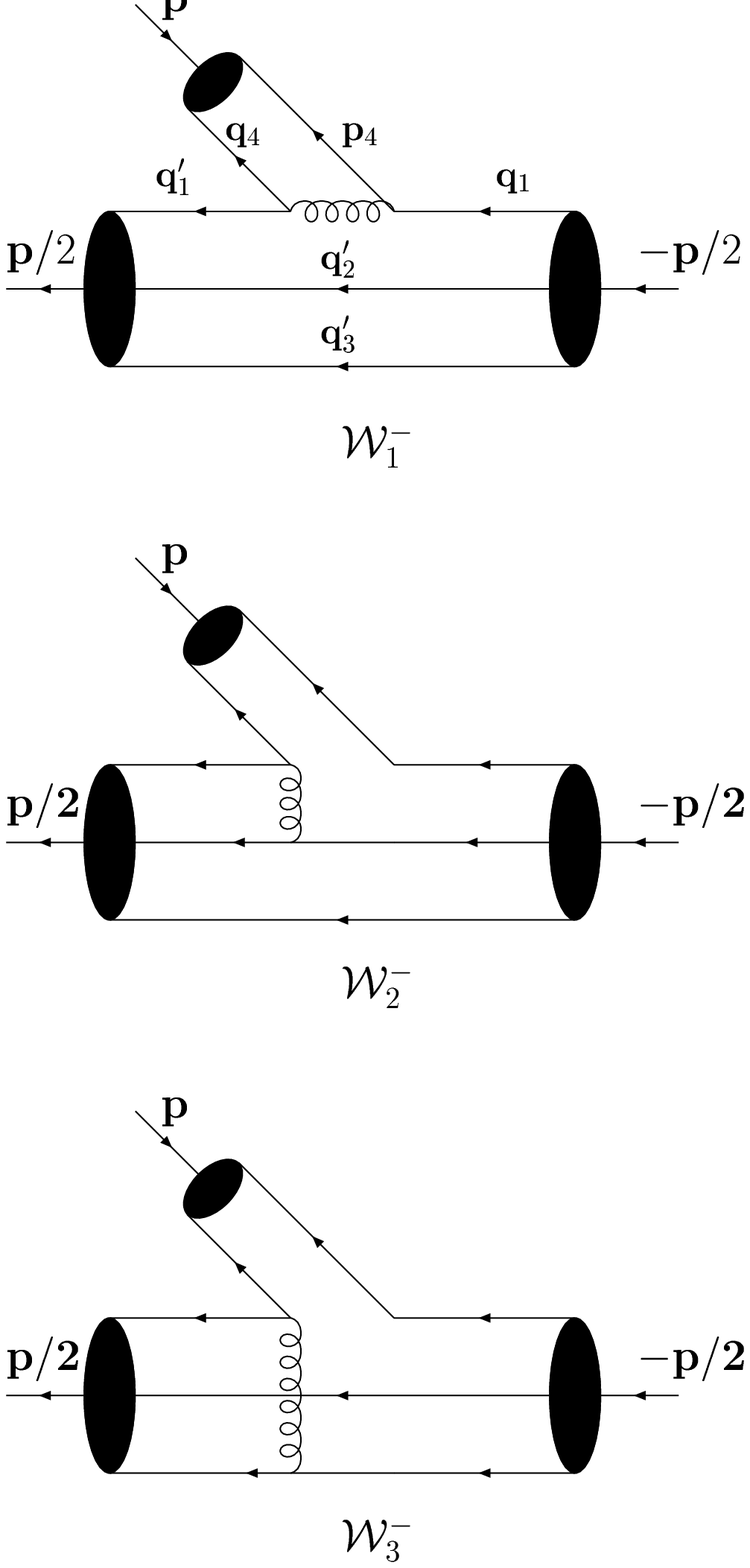}}
}
\vspace{-4.0cm}
\caption{Graphical representation of the functions ${\cal W}^{\pm}_{i},\;\;
i=1,2,3$.}
\label{W's}
\end{figure}

The pion-nucleon vertex function can be written generally as
\beq
W = \sum_{i=1}^{3}\left( W^+_i + W^-_i \right),
\label{Wis}
\eeq
where the $W^{\pm}_{i}$'s are of the general form (for simplicity we suppress
the color and spin-flavor wave functions in the following)
\beq
W^{\pm}_{i}(\bp,\bp';\bk) = \int \frac{d\bq\,d\bq'\,d\bq''}{(2\pi)^9} \, 
V_{\Gamma}(\bq-\bq') \,\Psi^*_{\bp'}(\bq'_1\bq'_2\bq'_3)\, 
{\cal W}^{\pm}_{i}[\Gamma, \phi_{\bf k}]\,
\Psi_{\bp}(\bq_1\bq_2\bq_3),
\label{Wpm12}
\eeq
where the ${\cal W}^{\pm}_{i}$'s involve the Dirac spinors and the pion wave
functions. In Figure~1 we present a pictorial representation of the different
contributions to the vertex function. Explicitly, the ${\cal W}^{\pm}_{i}$'s 
are given by
\bea
{\cal W}^{+}_1 &=& [u^{\dagger}(\bq'_1)T^a \Gamma u(\bq_4)]\,
\phi^+_{\bk}(\bp_4)\,[v^{\dagger}(-\bp_4)T^a \Gamma u(\bq_1)]\,,
\label{W+1} \\
{\cal W}^{+}_2 &=& \phi^+_{\bk}(\bp_4)\,[v^{\dagger}(-\bp_4)T^a \Gamma 
u(\bq_1)]\,[u^{\dagger}(\bq'_2)T^a \Gamma u(\bq_2)]\,,
\label{W+2} \\
{\cal W}^{+}_3 &=& {\cal W}^{+}_2\,,
\label{W+3} \\
{\cal W}^{-}_1 &=& [u^{\dagger}(\bq'_1)T^a \Gamma v(-\bq_4)]\,
\phi^{-T}_{\bk}(\bp_4)\,[u^{\dagger}(\bp_4)T^a \Gamma u(\bq_1)]\,,
\label{W-1} \\
{\cal W}^{-}_2 &=& [u^{\dagger}(\bq'_2)T^a \Gamma u(\bq_2)]\,
[u^{\dagger}(\bq'_1)T^a \Gamma v(-\bq_4)]\,\phi^{-T}_{\bk}(\bp_4)\,,
\label{W-2} \\
{\cal W}^{-}_3 &=& {\cal W}^{-}_2\,.
\label{W-3} 
\eea
In these formulas, the quark momenta in the initial (final) nucleon 
$\bq_1,\bq_2,\bq_3$ ($\bq'_1,\bq'_2,\bq'_3$) and the momenta of the quark 
and antiquark in the pion, ${\bp_4}$ and ${\bq_4}$, are expressed in terms 
of the loop momenta $\bq, \bq', \bq''$ by momentum conservation (see 
Fig.~\ref{W's}) . 

Once the effective baryon-pion Hamiltonian is obtained, one can calculate 
the pionic corrections to baryon properties as in the CBM and in the
traditional Chew-Low model. This will be done in the next section.

Before leaving this section, we recall that the use of the Breit frame is 
essential in calculations of form factors (vertex functions) in static 
models~\cite{JerryTony}, like the present one. This is true 
for composite models for which approximate solutions that maintain 
relativistic covariance are very difficult to implement. This was the
case for all old, static source, pion-nucleon models like the 
Chew-Low model~\cite{ChewLow}. In particular, as explained in 
Ref.~\cite{JerryTony}, electromagnetic gauge invariance is respected in 
this frame. Therefore, in calculating loop corrections to baryon properties, 
we employ the Breit-frame vertex functions. In the Breit frame, we denote 
the incoming pion and nucleon momenta by $\bp$ and $-\bp/2$, respectively, 
and the outgoing nucleon momentum by $\bp/2$. In this frame, the internal
momenta of quarks and antiquarks $\bq'_1,\bq'_2,\cdots$ are given in terms 
of the loop momenta $\bq, \bq'$ and $\bq''$ as 
\bea
\bq_1  &=& \bp/2 + \bq  + \bq''   \hspace{1.5cm}
\bq'_1 = \bp/2 + \bq' + \bq'' \nn\\
\bq_2  &=& \bq'_2 = - \bq' + \bq''   \hspace{1.25cm}
\bq_3  = \bq'_3 = - 2 \bq'' \nn\\
\bp_4  &=& \bp/2 - \bq  - \bq''  \hspace{1.5cm}
\bq_4  = \bp/2 - \bq  - \bq'' \,,
\label{momW+}
\eea
for the vertex ${\cal W}^+$ and
\bea
\bq_1  &=& - \bp/2 + \bq'  + \bq'' \hspace{1.5cm}
\bq'_1 = \bp/2 + \bq' + \bq'' \nn\\
\bq_2  &=& \bq'_2 = - \bq' + \bq'' \hspace{1.6cm}
\bq_3  = \bq'_3 = - 2 \bq'' \nn\\
\bp_4  &=& - \bp/2 + \bq  + \bq'' \hspace{1.5cm}
\bq_4  = - \bp/2 - \bq  - \bq'' \,,
\label{momW-}
\eea
for the vertex ${\cal W}^-$. In following equations, we also denote the 
vertex function as $W(-\bp/2,\bp/2;\bp) \equiv W(\bp)$. 

\section{Self-energy correction to baryon masses}
\label{sec:selfenergy}

For completeness we review the derivation of the expression for the 
self-energy correction from the effective baryon-pion Hamiltonian 
of Eq.~(\ref{partit}) in the ``one-pion-in-the air'' 
approximation~\cite{{ChewLow},{CBMorig}}. The baryon self-energy is defined 
as the difference of bare- and dressed-baryon energies:
\beq
\Sigma(E_B) = E_B - E^{(0)}_B .
\label{defSigma}
\eeq
The physical baryon mass $M_B$ is given by the (in general, nonlinear) 
equation
\beq
M_B = M^{(0)}_B + \Sigma(M_B)\,, 
\label{M_B}
\eeq
where $M^{(0)}_B$ is the bare baryon mass (i.e. without pionic corrections)
and $\Sigma(E_B)$ is the self-energy function. 

Let $|B \ran$ denote the physical baryon state, and $|B_0 \ran$ the ``bare''
undressed state. Let $Z^B_2$ be the probability of finding $|B_0 \ran$ in 
$|B \ran$. Then one can write
\beq
|B \ran = \sqrt{Z^B_2}\,|B_0 \ran + \Lambda |B \ran \,,
\label{B}
\eeq
where $\Lambda$ is a projection operator that projects out the component
$|B_0 \ran$ from $|B \ran$,
\beq
\Lambda = 1 - |B_0 \ran   \lan B_0|\,.
\label{Lambda}
\eeq
We have that 
\bea
\lan B| W |B_0 \ran &=& \lan B|(H - H_0)|B_0\ran = 
(E_B - E^{(0)}_B)\lan B|B_0 \ran  = \sqrt{Z^B_2}\; (E_B - E^{(0)}_B) \nn\\
& = & \sqrt{Z^B_2} \; \Sigma(E_B)\,.
\eea
We can now express $|B \ran$ in terms of $|B_0 \ran$ and the pion-baryon
interaction Hamiltonian $W$ as
\beq
|B \ran = \sqrt{Z^B_2}\,
\left[ 1 - \frac{1}{E_B - H_0 - \Lambda W \Lambda} \, W \right]
|B_0 \ran \,.
\label{BB_0}
\eeq
On the other hand, since
\beq
\lan B| W |B_0 \ran = \sqrt{Z^B_2}\, 
\lan B_0| W \, \frac{1}{ E_B - H_0 - \Lambda W \Lambda }\, W |B_0 \ran \,, 
\label{interm}
\eeq 
we have that the self-energy is given by
\bea
\Sigma(E_B) &=& \frac{1}{\sqrt{Z^B_2}} \, \lan B| W |B_0 \ran  \nn\\
&=& \lan B_0| W \, \frac{1}{ E_B - H_0 - \Lambda W \Lambda }\, 
W |B_0 \ran \,.
\label{Sigq-f}
\eea

This expression can be further approximated so as to avoid solving
complicated integral equations for the self-energy. We can manipulate the 
expression for $\Sigma$ to obtain (for details, see Ref.~\cite{CBMorig}): 
\beq
\Sigma(E_B) = \lan B_0| W \, \frac{1}{ E_B - H_0 - \Sigma_0 (E_B) } \, W 
|B_0 \ran \,,
\eeq
with
\beq
\Sigma_0(E_B) = W \Lambda \, \frac{1}{E_B - H_0} \, \Lambda W .
\label{Sig0}
\eeq
The approximation consists in absorbing $\Sigma_0(E_B)$ into $H_0$ such that
\beq
H_0 + \Sigma_0(E_B) \equiv \tilde{H_0}\,,
\label{H0tilde}
\eeq
with
\beq
\tilde{H_0} = \sum_{\alpha} E_{\alpha} \, B^{(0)\dag}_{\alpha}\, 
B^{(0)}_{\alpha} + \sum_j E_{\pi} \, M^{\dagger}_j \, M_j,
\label{H0-Hpiphys} 
\eeq
where $E_{\alpha}$ and $E_{\pi}$ are the {\em physical} energies. Therefore,
the baryon self-energy can be written as
\bea
\Sigma(E_B) = \langle B_0|W \frac{1}{E_B - \tilde H_0}W|B_0\rangle\,. 
\label{Sigma}
\eea
Finally, insertion of a sum over intermediate baryon-pion states in 
Eq.~(\ref{Sigma}) leads to
\beq
\Sigma(E_B) = \sum_n \langle B_0|W|n\rangle \frac{1}{E_B - E_n}
\langle n|W|B_0\rangle .
\label{sum}
\eeq

The structure vertex-propagator-vertex $W (E - \tilde H_0)^{-1}W$ 
in Eq.~(\ref{Sigma}) is an effective baryon-pion interaction. The main 
difference here with the hybrid  
approaches~\cite{{CT},{Bar},{BR},{Vento},{Jaffe},{CBMorig}} is that we do not 
have a point-like pion coupling to point-like quarks and antiquarks. The 
pion-baryon vertex arises through the ``Z-graphs'' in which the antiquark of 
the pion is annihilated with a quark of the ``initial'' baryon and the quark 
of the pion appears in the ``final" baryon. Therefore, the vertex function
incorporates not only the extension of the baryons, but also the extension of 
the pion.

We truncate the sum over the intermediate states in Eq.~(\ref{sum}) to the 
lowest mass states, namely the nucleon and the $\Delta(1232)$. In this case,
we obtain for the on-shell $N$ and $\Delta(1232)$ self-energies the coupled 
set of equations 
\bea
\Sigma_N(M_N) &=& \int \frac{d\bk}{(2\pi)^3} \, 
\left[ \frac{ |W_{N N}(\bk)|^2 } { M_N - (M_N + E_{\pi}(\bk))  } + 
\frac{ W_{N \Delta}(\bq)\, W_{\Delta N}(\bk) }
{ M_N - (M_\Delta + E_{\pi}(\bk)) } \right]\,,
\label{SigNuc} \\
\Sigma_\Delta(M_\Delta) &=& \int \frac{d\bk}{(2\pi)^3} \, 
\left[ \frac{ W_{\Delta N}(\bk) \, W_{N \Delta}(\bk) }
{ M_{\Delta}-(M_N + E_{\pi}(\bk)) }  + 
\frac{ |W_{\Delta \Delta}(\bk)|^2 }
{ M_\Delta-(M_{\Delta} + E_{\pi}(\bk)) } \right]\,.
\label{SigDel}
\eea
This is the final result for the pion loop correction for the nucleon
and $\Delta(1232)$. 

One important consequence of projecting the microscopic quark interaction 
onto hadronic states is that the leading nonanalytic (LNA) contributions in 
the pion mass as predicted by chiral perturbation theory are correctly 
obtained~\cite{TK}. In particular, as we discuss in the next section, 
Eq.~(\ref{SigNuc}) leads to an LNA contribution to the 
nucleon mass as predicted by QCD~\cite{LNA}, namely
\beq
M^{LNA}_N = - \frac{3}{ 16 \pi^2 f^2_{\pi} } g^2_A m^2_{\pi} \,.
\label{LNA}
\eeq
In models where the pion is treated as a point like particle, this result 
follows trivially~\cite{TK} from Eq.~(\ref{SigNuc}). In the context of the 
present model, where the pion is not treated covariantly, such a result does 
not follow in general for an arbitrary interaction. The difficulty is related 
to the fact the the pion dispersion relation, 
$E_\pi = \sqrt{\bk^2 + m^2_{\pi}}$, is not obtained in general in a
noncovariant model. In the CBM for example, the pion is point like and the 
normalization is correct from the very beginning. However, the microscopic 
quark interaction can be chosen such that the pion dispersion relation is 
correctly obtained~\cite{{LeYaou-mes},{Bic}}. These issues will be discussed 
in Section~\ref{sec:self-LNA}.

\section{The pion-nucleon and pion-$\Delta(1232)$ form factors}
\label{sec:numerical}

Our aim is to obtain an estimate for the numerical values of the 
pionic self-energies. It happens that nature has produced a
sort of low energy filter (chiral symmetry) for the details of strong
interactions. Indeed it is remarkable that although intermediate theoretical
concepts like gluon propagators, quark effective masses and so on, might
vary (in fact they are not gauge invariant and hence they are not physical
observables), chiral symmetry contrives for the final physical results, e.g.
hadronic masses and scattering lengths, to be largely insensitive to the above
mentioned theoretical uncertainties. The pion mass furnishes the ultimate
example: In the case of massless quarks, the pion mass is bound to be zero,
regardless of the form of the effective quark interaction {\em provided} it
supports the mechanism of spontaneous breakdown of chiral symmetry. 
The other example is provided by the $\pi -\pi $ scattering 
lengths~\cite{Weinberg} which  are equally independent of the form of the
quark kernel\cite{Ribeiro}. Furthermore it has become more and more 
evident through the accumulation of theoretical calculations on low energy 
hadronic phenomena, ranging from calculations on Euclidean space to 
instantaneous approximations and from harmonic kernels to linear confinement, 
that low energy hadronic phenomenology only seems to depend mildly on the 
details of the quark kernels used. To this extent, we will use for the 
quark-quark interaction a kernel of the form,
\bea
&& \Gamma     = \gamma^0\,, \label{Gamma} \\
&& V({\bf k}) = \frac{3}{4}(2 \pi )^{3}{K_0}^3 \Delta _k \delta ({\bf k}), 
\label{V}
\eea 
where $K_0$ is a free parameter. This potential has been widely used 
in the context of chiral symmetry breaking because it allows a great 
deal of simple analytic calculations (which is not the case for the linear 
potential). The harmonic potential basically differs from the linear 
potential in domains of the baryon-pion-baryon overlap kernel which contribute 
little to the total geometrical overlap so that, at least for results 
proportional to these overlaps, they should not differ too much.

The momentum-dependent part of the Salpeter amplitude for the baryon, 
$\Psi_{\bp}(\bq_1 \bq_2 \bq_3)$ in Eq.~(\ref{Bop}), is taken to be of a 
Gaussian form 
\beq
\Psi_{\bp}(\bq_1 \bq_2 \bq_3) = \frac{e^{-(\rho ^2+\lambda^2)/2 \alpha^2_B}} 
{{\cal N}_B (\bp)} ;\,,\hspace{1.0cm} 
\rho =\frac{{\bf p}_1-{\bf p}_2}{\sqrt{2}};\; 
\lambda =\frac{{\bf p}_1+{\bf p}_2-2 {\bf p}_3}{\sqrt{6}} ,
\label{baryon}
\end{equation}
where $\alpha^2_B$ is the variational parameter and ${\cal N}_B(\bp)$ is the 
normalization. Notice that since the integrations of the quark momenta in 
the functions $W^{\pm}_i$ in Eq.~(\ref{Wpm12}) are made through a Monte Carlo 
integration, the Gaussian ansatz is not essential and does not simplify our 
calculations, but we still use it to make contact with previous calculations. 

As in our previous calculation~\cite{fpiNN} for the pion-nucleon coupling 
constant, the Salpeter amplitudes $\phi^{\pm}_{\bk}(\bq)$ up to first 
order in $\bk$ are given by
\bea
\phi^+_{\bk}(\bq) &\simeq & { {\cal N}(\bk) }^{-1} \left[ + \sin \varphi(\bq) 
+ 
E_1(\bk) f_1(\bq) + 
i g_1(\bq) \, \bk\bcdot(\hat{\bq}\times \bsigma) \right] \chi_\pi 
{\cal S}_{color}, \label{eq.phi+} \\
\phi^-_{\bk}(\bq) &\simeq & { {\cal N}(\bk) }^{-1}  \left[ - \sin \varphi(\bq) 
+ E_1(\bk) f_1(\bq) - 
i g_1(\bq) \, \bk\bcdot(\hat{\bq}\times \bsigma) \right] \chi_\pi 
{\cal S}_{color}\label{eq.phi-} \,,
\eea
where $\varphi$ is the chiral angle and $E_1(\bk)$ is the first-order 
correction to the pion energy. The normalization ${\cal N}(\bk)$ is 
proportional to $E_1(\bk)$ and is given as
\beq
{\cal N}^2(\bk) = 4 E_1(\bk) \int \frac{d\bq}{(2\pi)^3} \, 
\sin\varphi(q)\, f_1(\bq) \equiv E_1 a^2 .
\label{Na}
\eeq
The energy $E_1(\bk)$ is given in terms of the second derivatives of the 
diagonal components of the Salpeter kernel with respect to $\bk$ and its 
explicit form is given in Eq.~(24) of Ref.~\cite{fpiNN}. Note that the 
truncation up to first order in $\bk$ of the Salpeter amplitude 
constitutes a reasonable approximation due to the fact that c.m. 
momenta-dependent distortions of the pion and nucleon wave functions are 
geometrically damped because of the geometric overlap kernel 
integrations for the functions $W^{\pm}_i$ in Eq.~(\ref{Wpm12}) --
see Ref.~\cite{emilio}. Explicit numerical solutions were obtained in 
Ref.~\cite{fpiNN} for the functions $f_1(\bq)$ and $g_1(\bq)$.  

For completeness, we initially repeat the results of Ref.~\cite{fpiNN} for 
the coupling constants $f_{\pi NN}$ and $f_{\pi N\Delta}$. 
In Ref.~\cite{fpiNN}, they were obtained as
\bea
\frac{f_{\pi NN}}{m_{\pi}} \, \bsigma_N\cdot\bp &=&
\frac{5}{3\sqrt{3}}\,\frac{{\cal O}_{fs}(p)}{2 a }\,\bsigma_N\cdot\hat{\bp} \,,
\label{O(p)}\\
\frac{f_{\pi N\Delta}}{m_{\pi}} \bS\cdot\bp &=& 
\left[\frac{2\sqrt{2}}{\sqrt{3}}
\frac{{\cal O}_{fs}(p)}{2 a } + \sqrt{2}\frac{{\cal O}'_{fs}(p)}{2 a}\right]
\bS\cdot\hat{\bp} \,,
\eea
where the isospin matrix is omitted and 
\beq
{\cal O}'_{fs}(p) \simeq 0 \,,\hspace{1.0cm}
{\cal O}_{fs} (p) = - \frac{\int [d\bq] \,(a^+ + a^- + b^+ + b^-)}
{\int [d\bq] \,\Psi^*_{in} \Psi_{out}} \,,
\eeq
where $[d\bq]$ means integration over $\bq, \bq'$ and $\bq''$ (see
Eq.~(\ref{Wpm12}) ) and the set of functions ${a^+ ,a^- ,b^+ ,b^-}$ is 
given by,
\bea
a^+ &=& \phi^+\, \Biggl\{
\varphi'(q_1) {\bsigma\cdot\bnabla} \Psi_{out} +
\left[ \varphi'(q_1) + \frac{ \cos\varphi(q_1) }{q_1}\right] 
{\bsigma\cdot\hat{\bq_1}}\times( \hat{\bq_1}\times\bnabla \Psi_{out} ) si
\Biggr\} \, \Psi_{in} \,, \\
a^- &=& \Psi_{out} \, \Biggl\{ \varphi'(q'_1) \bsigma\cdot\bnabla\Psi_{in} +
\left[\varphi'(q'_1) + \frac{\cos\varphi(q'_1)}{q'_1}\right] 
\bsigma\cdot\hat{\bq}'_1\times(\hat{\bq}'_1\times\bnabla \Psi_{in})
\Biggr\} \, \phi^{-} \,, \\ 
b^+ &=& \phi^+ \, \frac{1-\sin\varphi(q'_1)}{2 q'_1}\,
\Biggl\{2 \, \frac{\cos \varphi(p_1) }{p_1} \bsigma\cdot\hat{\bq}'_1 
+\left[ \varphi'(q_1) + \frac{\cos\varphi(q_1)}{q_1} \right] 
\bsigma \cdot \hat{\bq}_1 \times (\hat {\bq}_1 \times \hat {\bq}'_1 )
\Biggr\} \nn\\
&&\times \, \Psi_{out} \, \Psi_{in} \,, \\
b^- &=& \frac{1-\sin \varphi (q_1)}{2 q_1} \, \Biggl\{ 2 \,
\frac{\cos \varphi(q'_1))}{q'_1} \bsigma \cdot \hat {\bq}_1 
+ \left[\varphi'(q_1) + \frac{\cos \varphi(q'_1)}{q'_1}\right]
\bsigma \cdot\hat {\bq}'_1 \times (\hat {\bq}'_1 \times \hat {\bq}_1)\Biggr\}
\nn\\  
&& \times \, \Psi_{out} \, \Psi_{in} \, \phi^-\,.
\eea
Here,  $\Psi_{in,out}$ stand for the baryon {\em in} and {\em out} Salpeter 
amplitudes and $ \phi^{+,-}$ represent the pion Salpeter amplitudes.  

The baryon-pion coupling constants are obtained as the zero limit of the 
nucleon (or $\Delta$) momentum, $\bp \rightarrow 0$, of the above overlap
functions. For simplicity, we are defining the couplings at zero momentum, 
and not at the physical pion mass. In order to facilitate the integration, in 
Ref.~\cite{fpiNN} a Gaussian parameterization for the $(\cos \varphi(k))/k$ 
and $(1-\sin \varphi(k))/k^2$ was used. Here, since we need the vertex 
function for $\bp \neq 0$, we use a Monte Carlo integration to perform the
multi dimensional integral that gives the overlap function and use the full 
numerical solution for the gap function (not the Gaussian parameterization). 
We first checked the correctness of our Monte Carlo integration with 
the result of Ref.~\cite{fpiNN} for the special case of $\bp =0$ using the 
same Gaussian parameterization as was used there. This was done by calculating 
${\cal O}_{fs}({\bp})$ for ${\bp}=(0,0,p_z)$ and finding the limit of
${\cal O}={\cal O}_{fs}/p_z$ when $p_z \rightarrow 0$ to obtain 
$f_{\pi NN}$. 

As in Ref.~\cite{fpiNN} we have used $K_0=247$~MeV for the 
strength of the potential. The variational determination of $\alpha$ of
the baryon amplitude, Eq.~(\ref{baryon}), leads to $\alpha_N = 1.2 K_0$. For
the $\Delta(1232)$, the result is not much different and therefore we use
$\alpha_N = \alpha_\Delta$. 

Introducing the quantities
\beq
F_1 = \frac{5}{3\sqrt{12}} \,, \hspace{1.0cm} F_2 = \sqrt{\frac{2}{3}}\,,
\hspace{1.0cm} F_3 = \frac{1}{3 \sqrt{12}}\,,
\label{F's}
\eeq
we can summarize the couplings of the pion to the nucleon and 
$\Delta(1232)$ as follows:  
\beq
f_{\pi NN} = F_1 \,{\cal O}(0) \, \frac{m_{\pi}}{a}\,, \hspace{1.0cm}
f_{\pi N\Delta} = F_2 \,{\cal O}(0) \, \frac{m_{\pi}}{a}\,,\hspace{1.0cm}
f_{\pi \Delta\Delta} = F_3 \,{\cal O}(0) \, \frac{m_{\pi}}{a}\,.
\eeq
For the value of $K_0$ given above, we have $m_{\pi}/a=3.47$. 
The numerical values for the couplings are then
\begin{equation}
f_{\pi NN} = 1.19\,,\hspace{1.0cm} f_{\pi N\Delta} = 2.02 \,,
\hspace{1.0cm} f_{\pi \Delta\Delta} = 0.24 \,.
\end{equation}
The effect of the Gaussian parameterization can be assessed by comparing
with the corresponding numbers of Ref.~\cite{fpiNN}. For example, 
$f_{\pi NN} \simeq 1.0$ and $f_{\pi N\Delta} = 1.8$ in Ref.~\cite{fpiNN};
the effect of the parameterization is therefore of the order of 20$\%$.

\begin{figure}
\centerline{\epsfxsize=15.0cm\epsfbox{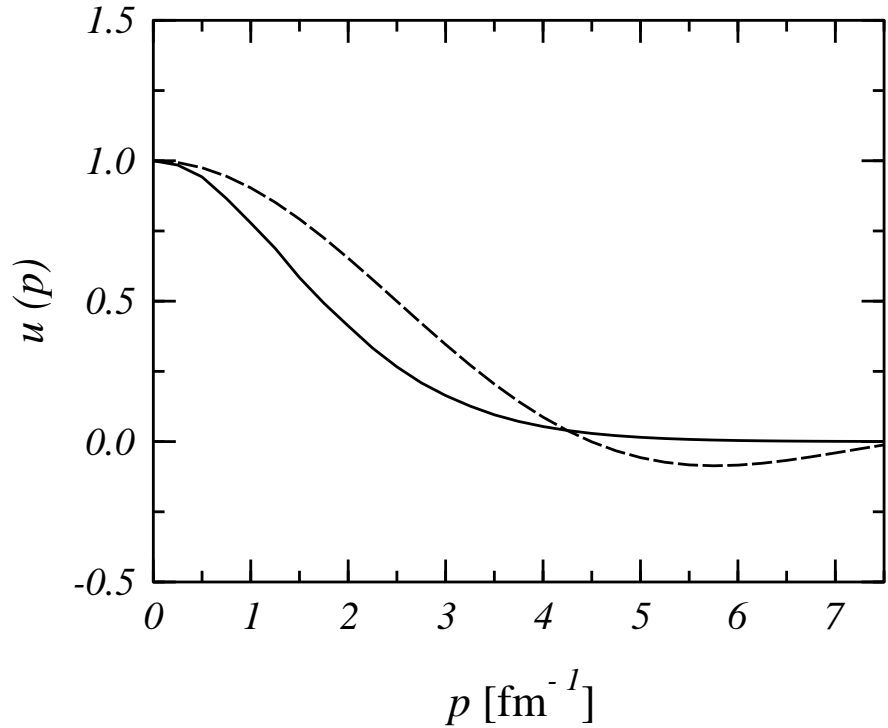}}
\vspace{0.5cm}
\caption{The function $u(p)$. The solid line is the form factor obtained with 
the baryon amplitude of Eq.~(\ref{baryon}) and pion Salpeter amplitudes of
Eqs.~(\ref{eq.phi+}) and (\ref{eq.phi-}). The dashed line is the CMB form 
factor of Eq.~(\ref{uCBM}) for $R=1$~fm.}
\label{u(p)}
\end{figure}

Next, we calculated the full overlap function for $\bp \neq 0$. In 
Figure~\ref{u(p)} we plot the function $u(p)={\cal O}(p)/{\cal O}(0)$ 
for the parameters given above. It is instructive to compare the momentum
dependence of this form factor with the one given by the 
CBM~\cite{{CBMorig},{CBMrev}}:
\beq
u(p) = 3 \, \frac{j_1(pR)}{pR}\,,
\label{uCBM}
\eeq
where $j_1$ is the spherical Bessel function and $R$ is the radius of the 
underlying MIT bag. The solid line is our result and the dashed one is 
the CBM result for $R=1$~fm. The faster falloff of our result is 
clearly a consequence of our Gaussian ansatz. As we will discuss soon, 
this rapid falloff will have the consequence of giving a smaller
value of the self-energy correction to the nucleon mass, as compared to
the corrections obtained with the CBM.

\section{Self-energy corrections to the nucleon and $\Delta(1232)$ masses}
\label{sec:self-LNA}

In this Section we present numerical results for the pionic
self-energy corrections to the nucleon and $\Delta(1232)$ masses
and discuss the LNA contribution to the masses. We start by
rewriting the vertex function in a manner to make clear the problem
with the pion dispersion relation. The pion energy is given, 
for low $\bk$, as~\cite{{LeYaou-mes},{Bic}}
\beq
E^2_1(\bk) = m^2_{\pi} + \bk^2 \sqrt{ \frac{
f^{(s)}_{\pi} }{f^{(t)}_{\pi}}  }\,,
\label{pidisp}
\eeq
where 
\beq
m^2_\pi = - \frac{ 2 m_q \lan\bar\psi\psi\ran }
{ \left( f^{(t)}_{\pi} \right)^2 } \,,\hspace{1.0cm}
\lan\bar\psi\psi\ran = - 6 \int \frac{d\bq}{(2\pi)^3}\, 
\sin \varphi(q)\,.
\label{mpi0}
\eeq
The point is that for an arbitrary quark-quark interaction one
obtains in general two different values for the pion decay constant, 
$f^{(t)}_{\pi}$ and $f^{(s)}_{\pi}$ (the explicit calculations can be 
found in Refs.~\cite{{LeYaou-mes},{Bic}}), depending on how one 
defines the decay constant. When using the time component of the axial 
current, one obtains $f^{(t)}_{\pi}$, and when using the space component 
one obtains $f^{(s)}_{\pi}$. However, as suggested in 
Ref.~\cite{LeYaou-mes}, and explicitly demonstrated in Ref.~\cite{Bic}, 
this problem can be cured by adding a transverse gluon interaction. 
Therefore, to illustrate the point of obtaining the correct LNA term from 
Eq.~(\ref{SigNuc}) with composite pions, we use the correct pion dispersion 
relation and assume $f^{(t)}_{\pi} = f^{(s)}_{\pi} \equiv f_{\pi}$ and 
denote $E_1(k) = \omega (k)$.

The normalization of the pion Salpeter amplitude, Eq.~(\ref{Na}),
can be rewritten as
\beq
{\cal N}^2(\bp) = 4 \omega(p) \int \frac{d\bk}{(2\pi)^3} \, 
\sin\varphi(q)\, f_1(\bq) = \frac{2}{3} \omega(p) f^2_{\pi} .
\label{Naf}
\eeq
That is, $a^2$ from Eq.~(\ref{Na}) is $2/3\,f^2_{\pi}$. We next extract 
from the vertex function (we concentrate on the NN form factor) this
normalization in the following way
\beq
W^i_{NN}(\bp) =\frac{1}{ \sqrt{2 \omega(p)} } \frac{G_A(p)}{2f_{\pi}}\,
\tau^i_N \, \bsigma_N \bcdot\bp \,. 
\label{Wour}
\eeq
The relation of the function $G_A(\bk^2)$ to the overlap function 
${\cal O}(p)$ can be trivially obtained by comparing with 
Eq.~(\ref{O(p)}). 

\begin{figure}
\centerline{\epsfxsize=10.0cm\epsfbox{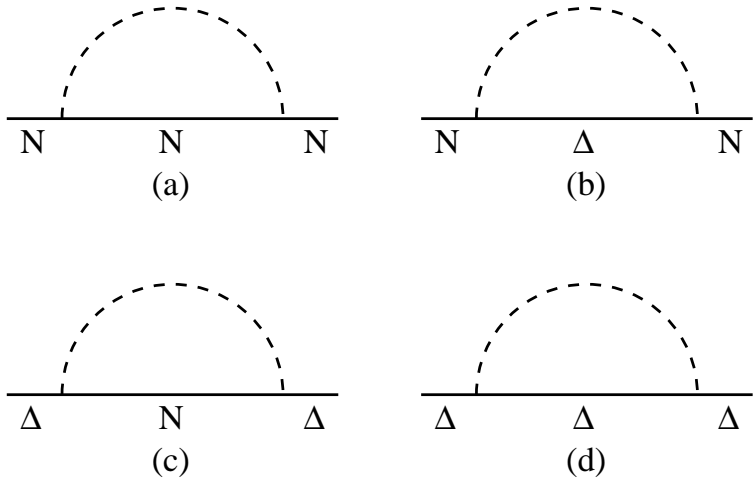}}
\vspace{0.5cm}
\caption{Schematic representation of the pion self-energy corrections to the 
nucleon ($N$) and delta ($\Delta$) masses.}
\label{self}
\end{figure}

Inserting Eq.~(\ref{Wour}) in the expression for the $N$ and $\Delta$ 
self-energies, Eqs.~(\ref{SigNuc}) and (\ref{SigDel}), and after performing
rather straightforward spin-isospin algebra one obtains 
\bea
M_N&=& M^{(0)}_N - f^2_0 \int_0^{\infty} dp\, \frac{p^4\,u^2(p)}{\omega^2(p)}
- \frac{32}{25} f^2_0 \, \int_0^{\infty} dp\, \frac{p^4\,
u^2(p)}{\omega(p)\left[\Delta M + \omega(p)\right]} \,,\label{MN} \\
M_\Delta &=& M^{(0)}_{\Delta} + \frac{8}{25}\ f^2_0 
\int_0^{\infty} dp\, \frac{p^4\,u^2(p)}
{\omega(p)\left[\Delta M - \omega(p)\right]} - f^2_0 \int_0^{\infty} dp\, 
\frac{p^4\,u^2(p)}{\omega^2(p)} \,,
\label{MD}
\eea
where 

\beq
\Delta M = M_{\Delta} - M_N\,,\hspace{1.0cm}
\omega(p)= \sqrt{ p^2+m^2_{\pi} },
\eeq
and 
\beq
u(p) = \frac{ {\cal O}(p) }{ {\cal O}(0) }\,,\hspace{1.0cm} 
f^2_0 = \frac{108}{m^2_{\pi}}\,\frac{f^2_{\pi NN}}{4\pi}.
\label{u_and_f0}
\eeq
%
%
Note that in principle we have different spatial dependencies for the $NN$,
$N\Delta$, $\cdots$ vertices, but for simplicity we have written them here as
being equal. A schematic representation of Eqs.~(\ref{MN}) and 
(\ref{MD}) is presented in Fig.~\ref{self}. It is important to note that 
these equations are not the ones one would obtain by simple perturbation 
theory; they are actually nonperturbative, because of the dependence on 
$\Delta M = M_{\Delta} - M_N$ on the r.h.s. 

It is easy now to obtain the LNA contributions to the masses~\cite{TK}. For 
the nucleon, the LNA contribution comes from the first term in Eq.~(\ref{MN}) 
by performing the integral. The integral can be done by transforming it into a 
contour integral and making use of Cauchy's theorem. The result is 
Eq.~(\ref{LNA}). For the $\Delta$, the LNA contribution follows in a similar 
way from the last term in Eq.~(\ref{MD}). 

To conclude, we discuss numerical results for the pionic corrections.
Initially we solve variationally the bare nucleon case. As discussed above, 
using $K_0=247$~MeV, we obtain for the variational size parameter the value 
$\alpha_N = 1.2 K_0$. We also use here $\alpha_N = \alpha_\Delta$. This 
leads to the following values for the bare $N$ and $\Delta$ masses 
\beq
M^{(0)}_N = 1174\;\text{MeV}\,,\hspace{1.0cm}
M^{(0)}_\Delta = 1373\;\text{MeV}\,.
\label{bare}
\eeq 
The difference between the masses, of the order of $200$~MeV, comes from the 
hyperfine splitting induced by the confining interaction. Given these values, 
we solve the two self-consistent equations given in Eqs.~(\ref{MN}) and 
(\ref{MD}). They are solved by iteration. We obtain for the masses
\beq
M_N = 1125 \;\text{MeV}\,,\hspace{1.0cm}
M_\Delta = 1342 \;\text{MeV}\,.
\label{dressed}
\eeq   
Comparing with the values above, we see that the pionic effect is relatively
small, as it should be, and of the order of $50$~MeV for the $N$
and $30$~MeV for the $\Delta$. The pionic effect is smaller for the 
$\Delta$, as one expects from spin-isospin considerations~\cite{TK}. The
results obtained with the CBM for a $R=1$~fm are a bit larger~\cite{CBMrev}. 
The difference can be traced to the rapid falloff of the form factor in our 
model.

We certainly do not expect these numbers to be definitive. Once more realistic
microscopic quark interactions and ansatze for the baryon wave function are 
used, they might be improved. However, independently of the microscopic model, 
our scheme is general and able to incorporate such interactions and new
baryon amplitudes. I would be of particular interest to have the numbers for a 
linear confining interaction with short range gluonic interactions that 
respect asymptotic freedom.

\section{Conclusions and future perspectives}
\label{sec:conclusion}

We developed a calculational scheme to calculate chiral loop corrections to 
properties of composite baryons with composite pions. The composite baryons 
and pions are bound states derived from a microscopic chiral quark model 
inspired in Coulomb gauge QCD and provides a generalization of the 
Nambu--Jona-Lasinio model to include confinement and asymptotic freedom. 
An effective chiral hadronic model is constructed by projecting the 
microscopic quark Hamiltonian onto a Fock-space basis of single composite 
hadronic states. The composite pions and baryons are obtained from the same 
microscopic Hamiltonian that describes the chiral vacuum condensate. The 
projection of the quark Hamiltonian onto the pion states is nontrivial 
because of the two-component nature of the Salpeter amplitude of the pion. 
As explained before, the two components correspond to positive and 
negative energies which complicates the Fock-space representation of the pion 
state. The projection is made possible by  rephrasing the formalism of the 
Salpeter equation in terms of the RPA equations. 

The development of models and calculational methods of the sort described
in the present paper are relevant in the context of a phenomenological
understanding of nonperturbative phenomena of strong QCD like confinement and 
dynamical chiral symmetry breaking. Eventually full lattice QCD simulations
aimed at studying hadronic structure will be available and phenomenological 
models will play a central role in the interpretation of the data generated. 
The developments of the present paper are of particular interest for the
first-principle developments based on the QCD Hamiltonian,  such as the 
nonperturbative renormalization program for the QCD Hamiltonian~\cite{renorm} 
and Hamiltonian lattice QCD~\cite{HLQCD}. We intend to implement the technique
developed here to such first-principle QCD calculations.

We illustrated the applicability of the formalism with a numerical calculation
using a simple microscopic interaction, namely a confining harmonic potential,
and a simple Gaussian ansatz for the baryon amplitude. This very same $S$-wave
interaction has been used  in a variety of earlier calculations, such as 
meson and baryon spectroscopy and $S$-wave nucleon-nucleon interaction. 
Numerical results were obtained here for the pion-nucleon form factor and 
for the pionic self-energy corrections to the nucleon and $\Delta(1232)$ 
masses in the nonperturbative one-loop approximation. Despite the simplicity 
of the interaction, the results obtained are very reasonable.

For the future, the most pressing development would be to use a microscopic
interaction that is consistent with asymptotic freedom and describes 
confinement by a linear potential. The calculation of the pion wave function 
beyond lowest order in momentum must be implemented and the variational ansatz
for the baryon amplitude must be improved. A more ambitious development would 
be to include explicit gluonic degrees of freedom. In this case  
renormalization issues will show up and the new techniques such as discussed 
in Ref.~\cite{renorm} will certainly be useful. Another very interesting
direction would be to employ the techniques developed here in Hamiltonian
lattice QCD.

\acknowledgements
This work was partially supported by CNPq (Brazil) and ICCTI (Portugal). 
The authors thank Nathan Berkovits for reading the manuscript and making 
suggestions on the presentation.

\vspace{1.0cm}

\end{document}